\documentclass[sigconf,screen]{acmart}
\definecolor{lightgreen}{rgb}{0.7, 0.9, 0.7}
\definecolor{lightred}{rgb}{1.0, 0.7, 0.7}
\usepackage{enumitem}
\usepackage[T1]{fontenc}
\usepackage[utf8]{inputenc}
\usepackage{geometry}
\usepackage{subcaption}  % 用于 (a) (b) 这样的子图
\usepackage{listings}
\AtBeginDocument{%
  }

\acmBooktitle{Companion Proceedings of the 33rd ACM Symposium on the Foundations of Software Engineering (FSE '25), June 23--27, 2025, Trondheim, Norway}
\acmConference[FSE Companion '25]{The 33rd ACM Joint European Software Engineering Conference and Symposium on the Foundations of Software Engineering}{June 23--27, 2025}{Trondheim, Norway}
\usepackage{balance}

\begin{document}
\acmYear{2025}\copyrightyear{2025}
\acmConference[FSE '25]{33rd ACM International Conference on the Foundations of Software Engineering}{June 23--28, 2025}{Trondheim, Norway}
\acmBooktitle{33rd ACM International Conference on the Foundations of Software Engineering (FSE '25), June 23--28, 2025, Trondheim, Norway}
%\acmDOI{10.1145/3696630.3728528}
%\acmISBN{979-8-4007-1276-0/25/06}
\title{Compiler Optimization Testing Based on Optimization-Guided Equivalence Transformations}

\author{Jingwen Wu}
\affiliation{%
  \institution{Shandong University}
  \city{}
  \state{}
  \country{}}
\email{elowen.jjw@gmail.com}

\author{Jiajing Zheng}
\affiliation{%
  \institution{Shandong University}
  \city{}
  \state{}
  \country{}}
\email{jiajing_zheng@163.com}

\author{Zhenyu Yang}
\affiliation{%
  \institution{Shandong University}
  \city{}
  \state{}
  \country{}}
\email{yangzycs@mail.sdu.edu.cn}

\author{Zhongxing Yu}
\authornote{Zhongxing Yu is the corresponding author.}
\affiliation{%
  \institution{Shandong University}
  \city{}
  \state{}
  \country{}}
\email{zhongxing.yu@sdu.edu.cn}

\begin{abstract}
Compiler optimization techniques are inherently complex, and rigorous testing of compiler optimization implementation is critical. Recent years have witnessed the emergence of testing approaches for uncovering incorrect optimization bugs, but these approaches rely heavily on the differential testing mechanism, which requires comparing outputs across multiple compilers. This dependency gives rise to important limitations, including that (1) the tested functionality must be consistently implemented across all compilers and (2) shared bugs remain undetected. Thus, false alarms can be produced and significant manual efforts will be required. To overcome the limitations, we propose a metamorphic testing approach inspired by compiler optimizations. The approach is driven by how to maximize compiler optimization opportunities while effectively judging optimization correctness. Specifically, our approach first employs tailored code construction strategies to generate input programs that satisfy optimization conditions, and then applies various compiler optimization transformations to create semantically equivalent test programs. By comparing the outputs of pre- and post-transformation programs, this approach effectively identifies incorrect optimization bugs. We conducted a preliminary evaluation of this approach on GCC and LLVM, and we have successfully detected five incorrect optimization bugs at the time of writing. This result demonstrates the effectiveness and potential of our approach.

\end{abstract}




\maketitle

\section{Introduction}
Compilers are essential for modern software development, translating high-level programming languages into machine code to enable efficient program execution. To address the growing complexity and diversity of software systems, compilers increasingly rely on advanced optimization techniques to improve performance and reduce resource consumption. However, these optimization techniques are inherently complex and prone to bugs \cite{alive,yang2011finding,compilerverification,rossopti}. For example, existing studies \cite{sun2016toward, zhou2021empirical} on two well-known compilers GCC and LLVM indicate that the optimization phase is more error-prone compared to other compiler passes. Optimization bugs can result in severe problems (\emph{e.g.} security vulnerabilities, performance degradation, and system instability), underscoring the urgent need for effective methods to detect and resolve these bugs. 

%Ensuring the correctness of compiler optimizations is, therefore, crucial for maintaining reliable software systems.

In the literature, testing \cite{yang2011finding,injectreal, rigoroustesting,emi,livecode,manycore,reduction,shader,multiple-fault}, translation validation \cite{alive,alive2,10.1007/BFb0054170,10.1145/349299.349314,10.1145/1993498.1993533,Stepp2011EqualityBasedTV}, and formal verification \cite{10.1145/1538788.1538814,10.1145/1111320.1111042,10.1145/2578855.2535866,compilerverification} are the three main categories of methods to improve the accuracy of the compiler. For its practicability and astonishing effectiveness, like quality assurance activities in other software systems \cite{dltesting,testweb,yujss,yuemse,YUgui,yuqsic}, testing remains the dominant technique. With regard to compiler optimization testing, existing studies focus primarily on detecting missed optimizations \cite{barany2018finding, zhang2023detection, liu2023exploring}, and few works are specifically tailored for incorrect optimizations that can cause a compile-time crash and produce incorrectly compiled code \cite{yang2011finding}. To the best of our knowledge, two works have been proposed to uncover incorrect optimizations during the past two years. Livinskii et al. \cite{livinskii2023fuzzing} redesign YARPGen \cite{livinskii2020random} with methods to improve loop code diversity, significantly increasing the likelihood of triggering optimizations and uncovering incorrect optimization bugs. Similarly, Xie et al. \cite{xie2024validating} introduce MopFuzzer, a fuzzing framework that maximizes runtime optimization interactions by encouraging multistage JVM optimizations. These works effectively generate test programs to expose incorrect optimization bugs, but they rely heavily on the differential testing mechanism \cite{differentialtesting,7958601}. Specifically, while crash errors can be directly revealed without requiring such testing, identifying silent wrong code errors depends on this testing, and such wrong code errors are common and are deemed the most harmful and difficult-to-detect compiler bugs \cite{injectreal}. The reliance on differential testing introduces two key limitations: (1) the functionality being tested must be consistently implemented across all compilers, and (2) shared bugs remain undetected if they present in every compiler. Thus, differential testing can lead to false alarms and requires extensive manual inspection to identify issues.

To overcome the limitations described above, we propose a new testing approach for incorrect optimizations based on the metamorphic testing mechanism \cite{mt}. The approach is inspired by compiler optimization change itself and is driven by how to maximize compiler optimization opportunities while effectively judging optimization correctness. Recognizing that the lack of customized input programs prevents certain optimizations from being triggered and thus their related bugs from being detected, we first develop code construction strategies to generate input programs that meet optimization requirements. We then apply various compiler optimization transformations (\emph{e.g.} loop optimization, data-flow optimization) to the generated programs in the previous step, creating semantically equivalent test programs. By comparing the outputs of pre- and post-transformation programs, our approach systematically identifies incorrect optimization bugs. As previous studies \cite{yang2011finding,alive} show that loop optimization is the buggiest optimization part, our implementation focuses on four loop optimization transforms up to now, including loop unrolling, loop-invariant code motion, loop unswitching, and loop fusion. We conducted a preliminary evaluation of this approach on GCC and LLVM, and we have successfully detected five incorrect optimization bugs at the time of writing. This result demonstrates the effectiveness and potential of our approach.

 The contributions of this paper are threefold:
(1) we propose a novel testing method for incorrect optimizations. (2) we instantiate this method with four specific loop optimization transforms. (3) we validate our approach on GCC and LLVM and identify five confirmed bugs, and our replication package is available at
\url{https://github.com/newolekcul/Optimization-testing}. 

\section{Approach Description}
% \vspace{-3pt}  % 减少与前文的距离
\begin{figure}[h]
\setlength{\abovecaptionskip}{5pt}  % 图标题与图之间的距离
\setlength{\belowcaptionskip}{-15pt} % 图标题与正文之间的距离
  \centering
  \includegraphics[width=0.8\linewidth]{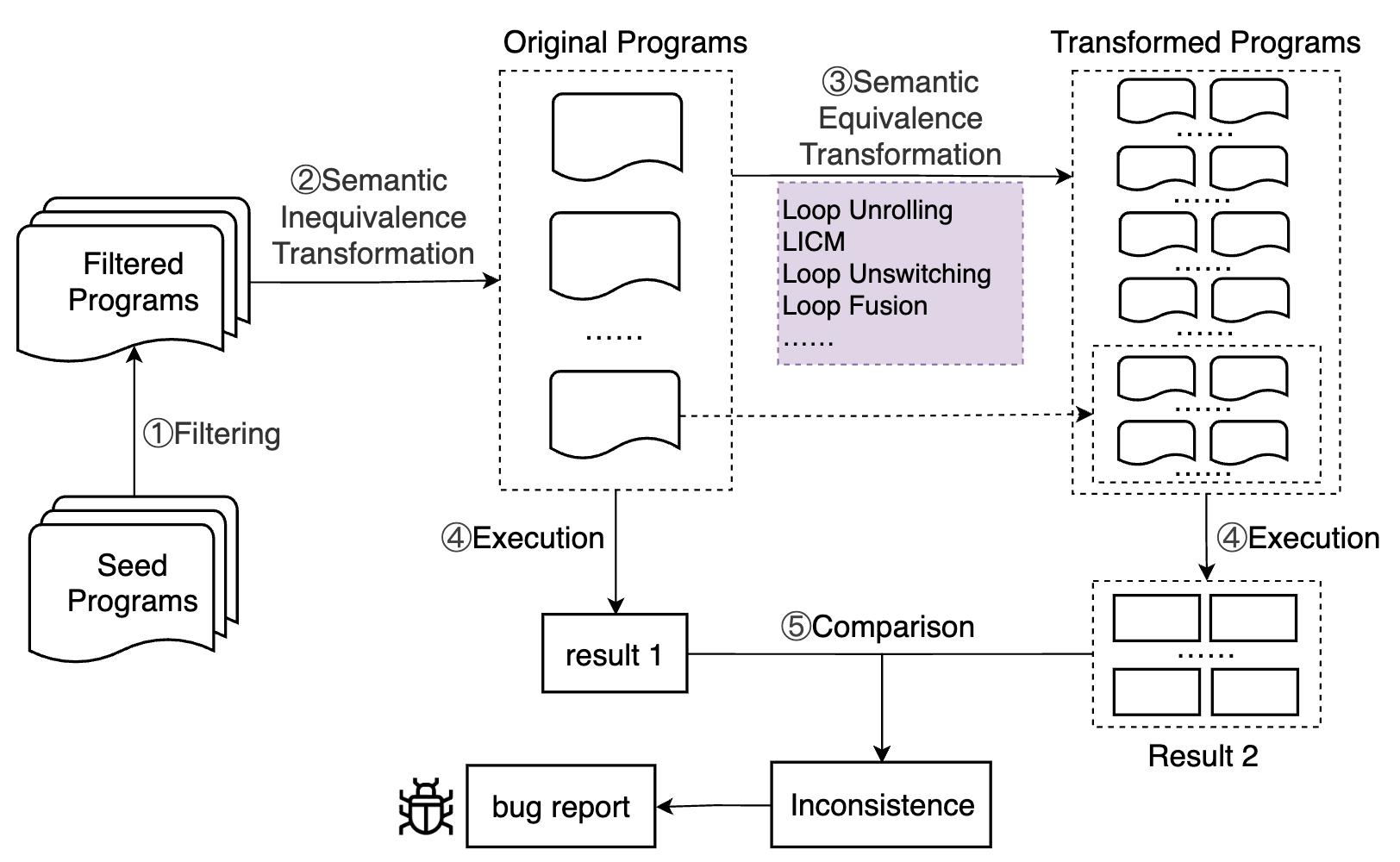}
  % \captionsetup{belowskip=-3em}
  \caption{Workflow of our approach}
  % \vspace{-0.6cm}
  \label{fig:framework} 
\end{figure}

\subsection{Overview}
Figure \ref{fig:framework} presents the workflow of the proposed testing method. It consists of the following five steps.

\noindent \textbf{\textit{ \underline{Step 1: Filtering. }}}
The seed programs need to be ensured to be compilable by the target compiler(s). To maintain reliable testing outcomes, seed programs that contain undefined behaviors (e.g., signed integer overflow or division by zero) or exhibit unpredictable behavior (e.g., caused by concurrency or randomness) are filtered out. The remaining programs are referred to as filtered programs.

\noindent \textbf{\textit{ \underline{Step 2: Semantic Inequivalence Transformation. }}}
Using filtered programs as input, we perform abstract syntax tree (AST) analysis to identify code segments suitable for transformation. Based on predefined loop optimization configurations, we construct test programs that satisfy the conditions for loop optimizations, referred to as original programs. In addition, certain construction processes require run-time information, such as loop iteration counts or the execution of specific statements. To gather this information, we instrument the filtered programs accordingly and execute them.

\noindent \textbf{\textit{ \underline{Step 3: Semantic Equivalence Transformation. }}}
Using original programs as input, we perform AST analysis to perform equivalence transformations based on the specific rules of the optimization method. Currently implemented optimizations include loop unrolling, loop-invariant code motion, loop unswitching, and loop fusion.

\noindent \textbf{\textit{ \underline{Step 4: Execution}}}
The original program and the transformed program are executed separately under consistent compiler settings and instruction configurations, producing their respective results.

\noindent \textbf{\textit{ \underline{Step 5: Comparison}}}
If the execution results of the original and transformed programs are consistent, no bug is identified; otherwise, a discrepancy indicates the presence of a potential bug.

\vspace{-2mm}
\subsection{Loop-based Equivalence Transformations}
We introduce four loop-based equivalence transformation methods designed to strategically modify or preserve program semantics. These transformations facilitate the activation of various loop optimizations, helping to uncover potential bugs in compiler optimization logic. The following paragraphs provide detailed descriptions.

\begin{figure*}[ht]
    % \vspace{8pt}
    \setlength{\abovecaptionskip}{5pt}  % 图标题与图之间的距离
    \setlength{\belowcaptionskip}{-8pt} % 图标题与正文之间的距离
    \centering
    \includegraphics[width=0.80\linewidth]{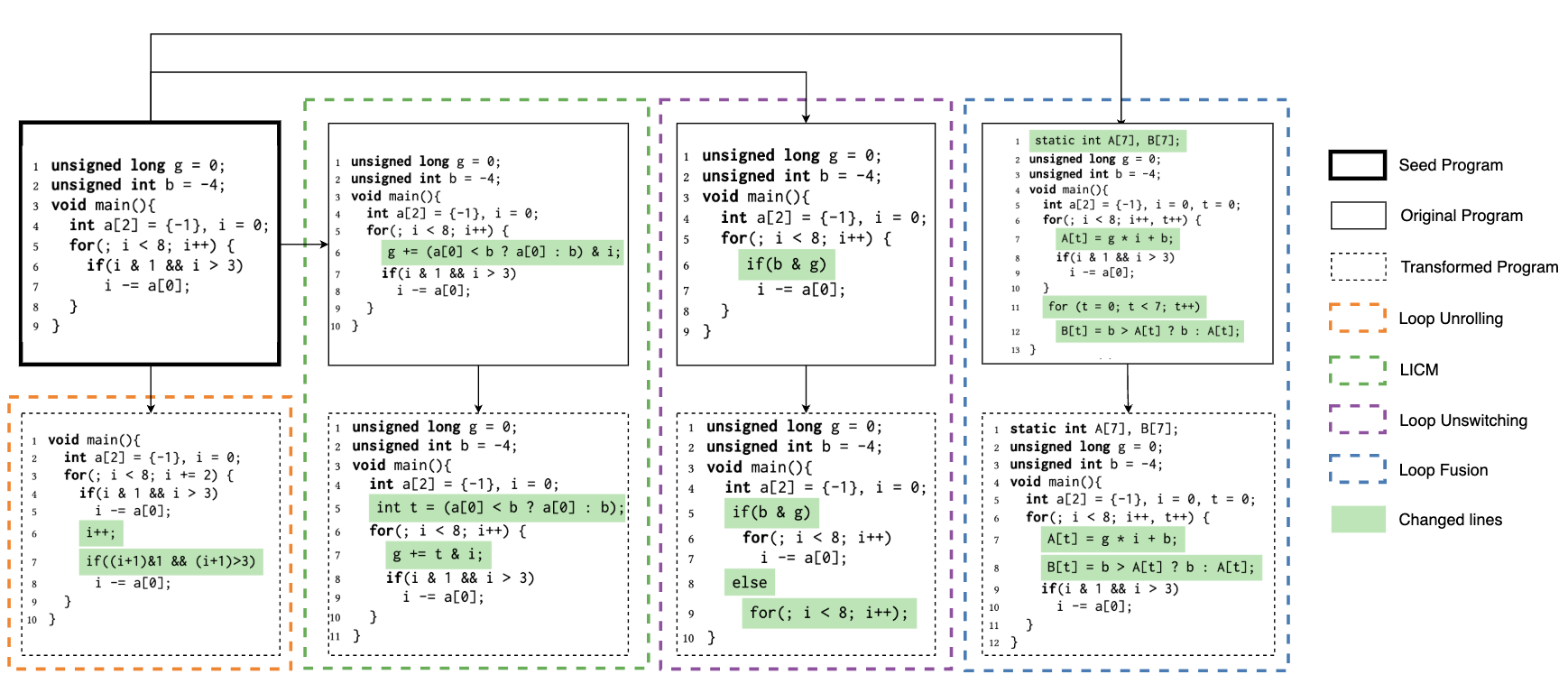}
    \caption{Examples of loop optimization transformation methods.}
    \label{fig:trans_example}
\end{figure*}

% \vspace{0.5em}
\textbf{\textit{Loop Unrolling. }} This transformation reduces the number of iterations by replicating the loop body multiple times per iteration, thereby minimizing overhead associated with operations such as counter increments and conditional checks. This method applies to any loop structure that satisfies the optimization conditions, allowing the seed program to serve directly as the original program without additional code construction. 

To implement loop unrolling, we begin with a standard loop structure. A critical aspect of loop unrolling is determining the unrolling factor, denoted as \texttt{k}, which specifies the number of loop body iterations executed in each cycle. This choice presents a trade-off: larger values of \texttt{k} reduce loop control overhead, but excessive code duplication can negatively impact cache performance. In this study, we determine the value of \texttt{k} by first calculating the total number of loop iterations, \texttt{n}, and then identify the factors of \texttt{n}. Since the exact value of \texttt{n} is unknown at compile time, each factor of \texttt{n} is tested sequentially as a potential \texttt{k}. For each unrolling step, operations applied to the loop index (e.g., increment or decrement) must be added after the unrolled iterations, except for the final unrolling step, where such operations are inherently excluded due to the loop's syntactic structure. Furthermore, if \texttt{n} is a prime number, the loop is divided into two parts: a main loop and a boundary section. The largest composite number \texttt{m} less than \texttt{n} is identified first. The main loop, with \texttt{m} iterations, is unrolled using the general method aforementioned, while the remaining \texttt{n-m} iterations are preserved in the boundary section without unrolling. 

% \vspace{0.5em}
 \textbf{\textit{Loop-Invariant Code Motion (LICM).}}
This transformation identifies instructions whose results remain unchanged across all iterations of a loop and relocates them outside the loop. 

To ensure semantic equivalence before and after this optimization, the following three conditions should be met: (1) The hoisted code should not rely on variables or states that change during loop iterations, ensuring that its result remains constant. (2) The hoisted instructions should not introduce side effects on loop control variables or external states (e.g., global variables, references, or pointer targets). (3) The reordering should not disrupt the sequence of dependent operations within the loop, thereby avoiding logic errors. 

For scenarios where a program lacks sufficient loop-invariant code, we can construct it artificially. Starting from the seed program, we first identify the variables that remain constant in all iterations of the loop. Using these variables, we construct an expression $E$ by combining them with various operators (e.g., arithmetic, bitwise, logical). The new expression $E$ is then inserted into any appropriate line within the loop, generating the original programs. Next, we apply the LICM rules to the newly introduced statement, relocating $E$ outside the loop to produce transformed programs. 

% \vspace{0.5em}
 \textbf{\textit{Loop Unswitching.}}
This transformation moves conditional statements from inside a loop to outside, transforming the loop by explicitly separating the conditional logic into distinct branches. Each branch contains its own loop version, corresponding to the respective outcomes of the conditional statement (e.g., the \texttt{if} and \texttt{else} branches).

To ensure correctness, the technique requires that the conditional expression be independent of loop variables and free of side effects. If the seed program does not meet these criteria, we modify the condition to include only loop-invariant expressions, constructed in a manner similar to those used in LICM, to generate the original programs. Next, we apply the loop unswitching rules to the original programs, generating the transformed programs. 

% \vspace{0.5em}
 \textbf{\textit{Loop Fusion.}}
This transformation combines multiple loops with the same iteration space into a single loop to improve data locality, cache utilization, and execution speed. This optimization is classified according to the data dependency between loop bodies: (1) independent scenarios where loop results do not depend on each other, allowing reordering without affecting correctness, and (2) dependent scenarios where one loop depends on the results of another, requiring preservation of execution order. In this study, our focus is on scenarios where the two loops exhibit data dependencies, as such cases provide greater opportunities to uncover incorrect optimization bugs. 

To ensure correctness, all fused loops should share the same index range and step size. These constraints guarantee logical consistency while enabling the performance advantages of loop fusion. In many cases, seed programs may not offer enough loops that meet these criteria. Therefore, we begin with the seed program and apply the following procedure to construct the original programs. First, we identify a loop and dynamically determine its iteration count during execution. Next, we construct a second loop with the same iteration count. In each loop, we introduce an array and assign it to the array with crafted right-hand expression. To simulate real-world dependencies, the second loop must include variables influenced by the first loop. After generating the original programs, we fuse the two loops into a single loop according to the loop fusion rules, producing the transformed programs.

For illustration purposes, Figure \ref{fig:trans_example} shows the four transformations that start from the seed program, then to the generation of the original program, and finally to the generation of transformed program. 

% \vspace{0.5em}
\section{Preliminary Evaluation}
\subsection{Evaluation Setup}
\noindent\textbf{\textit{Environment Setup.}}
We evaluate the latest versions of GCC and LLVM (at the time of our work) using our approach in the following configurations: Ubuntu 22.04 equipped with an Intel i7-10700F 2.90GHz 16-core CPU and 32GB of memory. Our evaluation covers five optimization levels (\texttt{-O0}, \texttt{-O1}, \texttt{-O2}, \texttt{-O3}, and \texttt{-Os}) as well as randomized combinations of optimization flags.

\noindent\textbf{\textit{Seed Programs.}}
We employ Csmith \cite{yang2011finding} to generate seed programs due to its ability to produce random, reliable, and highly customizable test cases. 
%Using a pseudo-random number generator, Csmith first selects language constructs such as variables, control structures, and function calls, and then combines them into complex and diverse test programs. 
To ensure the semantic validity of the generated programs, Csmith integrates internal consistency checks to eliminate undefined behavior. 
% This process guarantees that the test programs are both syntactically and semantically correct, thus enhancing the reliability of the testing process.
Additionally, Csmith offers extensive customization options (\emph{e.g.}, parameters for program size and data types), facilitating the targeted generation of the test case tailored for specific optimization scenarios and testing requirements.

\subsection{Results and Bug Analysis}
In our preliminary experiments, we identified five bugs that are confirmed by the developers at the time of writing. Specifically, we identified three bugs in GCC and two bugs in LLVM. These issues were triggered by code segments constructed using the loop-based equivalence transformations proposed in this study. The following paragraphs provide detailed analyses of representative cases. 

\begin{figure}[ht]
    % \vspace{8pt}
    \setlength{\abovecaptionskip}{3pt}  % 图标题与图之间的距离
    \setlength{\belowcaptionskip}{-5pt} % 图标题与正文之间的距离
    \centering
    \includegraphics[width=0.80\linewidth]{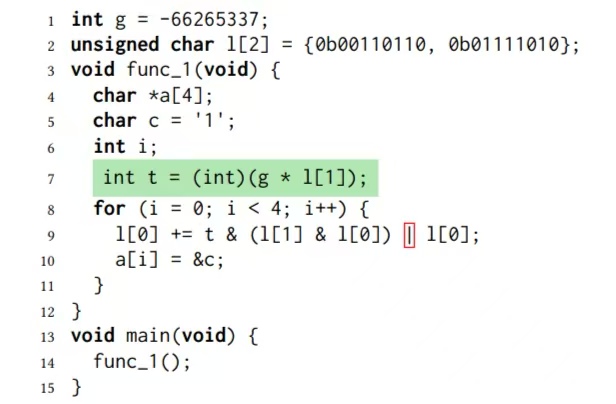}
    \caption{A sample that triggers a GCC compiler bug.\protect\footnotemark}
    \label{fig:bug_example1}
\end{figure}
\footnotetext{\url{https://gcc.gnu.org/bugzilla/show_bug.cgi?id=113669}}

Figure \ref{fig:bug_example1} shows a GCC bug that can be triggered by the loop-invariant code motion (LICM) transform.  
% the core of this bug lies in whether \texttt{t} (i.e., \texttt{(int)(g * l[1])}) is extracted outside the loop and how the compiler handles the multiplication overflow check. 
For the pre-transformation program, the statement \texttt{int t = (int)(g * l[1])} stays inside the loop. In this case, the computation of \texttt{t = (int)(g * l[1])} and its associated overflow detection occur for each loop iteration, allowing the compiled code to detect signed integer overflow at runtime and issue a warning accordingly. However, after the LICM transformation, the statement \texttt{int t = (int)(g * l[1])} is moved outside the loop and the code behavior changes under the \texttt{-O2} optimization level. In this case, the compiler analyzes the remaining loop statement \texttt{l[0] += t \& (l[1] \& l[0]) | l[0]} and determines that the higher-order bits are ``unnecessary'', leading it to erroneously simplify the operation to \texttt{l[0] += l[0]}. This simplification effectively removes the multiplication overflow check, preventing any signed integer overflow warning under \texttt{-O2} optimization level. This discrepancy reveals that the compiler incorrectly simplifies the multiplication and logical operations, eliminating the overflow detection mechanism.

\begin{figure}[ht]
    % \vspace{8pt}
    \setlength{\abovecaptionskip}{3pt}  % 图标题与图之间的距离
    \setlength{\belowcaptionskip}{-5pt} % 图标题与正文之间的距离
    \centering
    \includegraphics[width=0.80\linewidth]{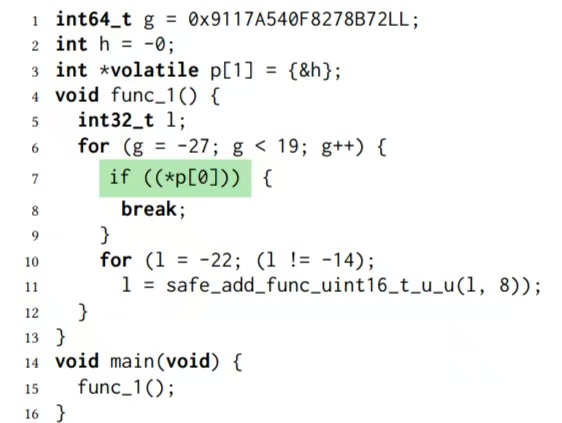}
    \caption{A sample that triggers a LLVM compiler bug.\protect\footnotemark}
    \label{fig:bug_example2}
\end{figure}

Figure \ref{fig:bug_example2} shows a LLVM bug triggered by the loop unswitching transformation. Notably, this bug exposes a critical flaw in LLVM’s optimization pipeline: the failure to preserve semantic equivalence between the pre- and post-transformation programs, particularly when handling infinite loops. For the pre-transformation program, the termination condition \texttt{l == -14} is unsatisfiable due to the constraints imposed by the \texttt{safe\_add\_func\_uint16\_t\_u\_u} function (which ensures that the addition result stays within the \texttt{uint16\_t} range), preventing the loop from terminating. However, LLVM’s speculative execution and dead-code elimination optimizations misinterpret the termination logic as redundant, causing it to bypass the infinite loop and leading to premature termination. After the loop unswitching transformation, the conditional logic is split into two branches, each containing its own version of the loop. Subsequent optimizations, such as constant folding, branch pruning, and loop peeling, further simplify the control flow, eliminating one branch due to assumed invariants. This results in a different loop structure and behavior compared to the pre-transformation program. The intended performance optimizations inadvertently introduce logical errors, causing the program to behave incorrectly and resulting in output inconsistencies, especially under high optimization levels.

\footnotetext{\url{https://github.com/llvm/llvm-project/issues/75809}}

\section{Related Work}
\noindent \textbf{Detecting Missed Compiler Optimizations.} Existing studies about compiler optimization testing primarily focus on
detecting missed optimizations \cite{barany2018finding,zhang2023detection,liu2023exploring,10.1145/3276496,10.1145/3503222.3507764}. Theodoridis et al. \cite{10.1145/3503222.3507764} propose an approach to detect missed optimizations by analyzing the live and dead basic blocks. Zhang et al. \cite{zhang2023detection} introduce MOD, a general method that detects missed optimizations through a manually curated mapping between different optimization phases. Liu et al. \cite{liu2023exploring} develop DITWO, a differential testing framework designed to uncover missed optimizations in WebAssembly optimizers.
% by comparing the performance of native x86 and WebAssembly executables. 

\noindent \textbf{Detecting Incorrect Compiler Optimizations.}
Two works have been proposed to uncover incorrect optimizations during the past two years. Livinskii et al. \cite{livinskii2023fuzzing} redesign YARPGen \cite{livinskii2020random} with methods to enhance loop code diversity, significantly increasing the likelihood of triggering optimizations. Similarly, Xie et al. \cite{xie2024validating} introduce MopFuzzer, a fuzzing framework that maximizes runtime optimization interactions by encouraging multi-stage JVM optimizations. Despite these advancements, the use of metamorphic testing \cite{chen2020metamorphic} methods for compiler testing remains underexplored.

\section{Conclusion}
In this paper, we propose a metamorphic testing approach inspired by compiler optimizations to identify incorrect optimization bugs. In particular, our approach first employs tailored code construction strategies to generate input programs that satisfy optimization conditions, and then applies various compiler optimization transformations to create semantically equivalent test programs. By comparing the outputs of pre- and post-transformation programs, this approach effectively identifies incorrect optimization bugs. Our current implementation focuses on four loop optimization transforms, and a preliminary evaluation on GCC and LLVM has successfully detected five incorrect optimization bugs at the time of writing.

% We develop strategies to generate input programs that align with optimization requirements. By applying equivalence transformations based on loop optimizations, we generate semantically equivalent programs and compare their outputs to detect discrepancies that may indicate bugs. We validate our approach through a preliminary evaluation on two open-source compilers GCC and LLVM, and we have successfully detected five incorrect optimization bugs at the time of our writing. In the future, we plan to do more thorough testing by incorporating additional test programs and exploring other compiler optimization transforms to further refine our approach.

\section*{Acknowledgments}
\noindent
We appreciate the reviewers for their insightful comments. This work was supported by National Natural Science Foundation of China (Grant No. 62102233), Shandong Province Overseas Outstanding Youth Fund (Grant No. 2022HWYQ-043), Joint Key Funds of National Natural Science Foundation of China (Grant No. U24A20244), and Qilu Young Scholar Program of Shandong University. 

% \section{Acknowledgments}
% Identification of funding sources and other support, and thanks to individuals and groups that assisted in the research.

%\begin{acks}
%To Robert, for the bagels and explaining CMYK and color spaces.
%\end{acks}


\bibliographystyle{ACM-Reference-Format}
\bibliography{sample-base}

\end{document}